\documentclass[aps,pre,twocolumn,showpacs,superscriptaddress,amsmath,amssymb,floatfix]{revtex4}
\usepackage[pdftex]{graphicx}

\newcommand{\flabel}[1]{\label{f:#1}}
\newcommand{\tlabel}[1]{\label{t:#1}}
\newcommand{\fig}[1]{Fig.~\ref{f:#1}}
\newcommand{\tab}[1]{Table~\ref{t:#1}}
\renewcommand{\vec}[1]{\mathbf{#1}}
\newcommand{\mean}[1]{\left\langle #1 \right\rangle}

\begin{document}

\title{Hard-disk equation of state: First-order liquid--hexatic
  transition in two dimensions with three simulation methods}

\author{Michael Engel}
\affiliation{Department of Chemical Engineering, University of Michigan, Ann Arbor, Michigan 48109, USA}

\author{Joshua A. Anderson}
\affiliation{Department of Chemical Engineering, University of Michigan, Ann Arbor, Michigan 48109, USA}

\author{Sharon C. Glotzer}
\email{sglotzer@umich.edu}
\affiliation{Department of Chemical Engineering, University of Michigan, Ann Arbor, Michigan 48109, USA}

\author{Masaharu Isobe}
\affiliation{Graduate School of Engineering, Nagoya Institute of Technology, Nagoya, 466-8555, Japan} 
\affiliation{Department of Chemistry, University of California, Berkeley, California 94720, USA}

\author{Etienne P. Bernard} 
\affiliation{Department of Physics, Massachusetts Institute of Technology, Cambridge, Massachusetts 02139, USA}

\author{Werner Krauth}
\email{werner.krauth@ens.fr}
\affiliation{Laboratoire de Physique Statistique, \'{E}cole Normale Sup\'{e}rieure, UPMC, CNRS, 24 Rue Lhomond, 75231 Paris Cedex 05, France}

\date{\today}

\begin{abstract}
We report large-scale computer simulations of the hard-disk system at
high densities in the region of the melting transition.  Our
simulations reproduce the equation of state, previously obtained using
the event-chain Monte Carlo algorithm, with a massively parallel
implementation of the local Monte Carlo method and with event-driven
molecular dynamics.  We analyze the relative performance of these
simulation methods to sample configuration space and approach
equilibrium.  Our results confirm the first-order nature of the
melting phase transition in hard disks.  Phase coexistence is
visualized for individual configurations via the orientational order
parameter field. The analysis of positional order confirms the
existence of the hexatic phase.
\end{abstract} 


\maketitle 

\section{Introduction} 
The phase behavior of hard disks is one of the oldest and most studied
problems in computational statistical mechanics. It inspired the use of Markov-chain
Monte Carlo~\cite{Metropolis_1953} as well as molecular
dynamics~\cite{Alder_1957}. Important progress in understanding
hard-disk melting~\cite{Alder_1962,Strandburg_1988,Dash_1999} was made
recently.  Using the event-chain Monte Carlo algorithm
(ECMC)~\cite{Bernard_2009}, a first-order liquid--hexatic transition
was identified~\cite{Bernard_2011}.  This transition from the
low-density phase to an intermediate phase precedes a continuous
hexatic--solid transition, and thus the liquid transforms to a solid
through an intermediate hexatic phase.

Controversy concerning the nature of hard-disk melting has persisted
for decades.  Indeed, the recently discovered first-order
liquid--hexatic melting transition differs from the standard KTHNY
scenario~\cite{Kosterlitz_1972,Halperin_1978,Nelson_1979,
  Young_1979,Jaster_1999,Mak_2006}, which predicts continuous
transitions both from the liquid to the hexatic and from the hexatic
to the solid. It is also at variance with the first-order
liquid--solid transition scenario, which exhibits no intermediate
hexatic phase and has been much
discussed~\cite{Alder_1962,Fisher_1979,Chui_1983,
  Kleinert_1983,Ramakrishnan_1982,Weber_1995, Mak_2006}.  Near the
critical density, the system is correlated across roughly a hundred
disk radii, and the hard-disk liquid--hexatic transition is thus
weakly first-order~\cite{Bernard_2011}, with only a small
discontinuity in density at the transition.

For several decades, the algorithms used for this
problem~\cite{Zollweg_1992,Lee_1992,Weber_1995,Jaster_1999,Mak_2006}
were unable to equilibrate systems sufficiently larger than the
spatial correlation length to reliably investigate the existence and
the nature of the hexatic phase.  This was one origin for the
controversy surrounding this problem.  Another reason was that the
manifestation of a first-order transition in the $NVT$ ensemble, and
in particular the fundamental difference between a van der Waals loop
and a Mayer-Wood loop indicating equilibrium phase coexistence, was
not universally accepted in the hard-disk community, although it had
been clearly discussed in the
literature~\cite{Mayer_1965,Furukawa_1982,Schrader_2009}.

Here, we complement and compare the recent event-chain results with a
massively parallel implementation of the local Monte Carlo algorithm
(MPMC)~\cite{Anderson_2012}, and with event-driven molecular dynamics
(EDMD)~\cite{Isobe_1999}. These methods provide us with by far the
largest independent data sets ever acquired for the hard-disk melting
transition. Our simulations reproduce to very high precision the
equation of state of Ref.~\cite{Bernard_2011}, illustrating phase
separation. To characterize the nature of the two hard-disk phase
transitions, we graphically represent the orientational and positional
order parameter fields and analyze positional correlation functions.

\section{Simulation methods}

\subsection{System definition}
We consider a system of $N$ hard disks of radius $\sigma$ in a square
box of size $L\times L$. The phase diagram of the system depends only
on the density (packing fraction) $\eta = N\pi \sigma^2 / L^2$, as the
pressure is proportional to the temperature $T$. The dimensionless
pressure is given by
\begin{equation}
  P^* = \frac{(2\sigma)^2}{m\langle v_x^2\rangle } P= \beta P(2\sigma)^2
\end{equation}
with the inverse temperature $\beta$, mass $m$, and the velocity along
one axis $v_x$. All simulations are conducted in the $NVT$ ensemble
and, although our algorithms differ in the way they evolve the system,
they all sample the same equilibrium probability distribution in
configuration space. At finite $N$, the equilibrium phase coexistence
that we will observe is specific to this ensemble and absent, for
example, in the $NPT$ ensemble. As for any model with short-range
interactions, the thermodynamic limit is independent of the ensemble.

\subsection{Algorithms and implementations}

Local Monte Carlo (LMC) has been a popular simulation method for hard
disks~\cite{Metropolis_1953,Lee_1992,Weber_1995,Mak_2006}.  At each
time step, one random disk is selected and a trial move is applied to
it.  The move is accepted unless it results in an overlap with another
disk. LMC is both relatively inefficient in sampling configuration
space and inherently serial, limiting the size for which the system
can be brought to equilibrium at high densities $\eta \sim 0.7$ to
about $N \sim 10^5$ particles (see \cite{Krauth_2006} for a basic
discussion). Our alternative approaches utilize modern computer
resources more efficiently and equilibrate the system faster. They
also provide independent checks of the equilibrium phase behavior. LMC
is used for comparison and as a reference to previous work.

Massively parallel Monte Carlo (MPMC)~\cite{Anderson_2012} is a
parallel extension of LMC.  It again applies a local trial move, but
maximizes the number of simultaneous updates. MPMC extends the stripe
decomposition method~\cite{Uhlherr_2002} to a massive number of
threads using a four-color checkerboard scheme~\cite{Pawley_1985}. By
placing disks into cells of width $w \gtrsim 2 \sigma$, concurrent
threads execute over one out of four subsets of cells in parallel.
Within each cell, particles are chosen for trial moves in a shuffled
order.  The number of trial moves is fixed independent of cell
occupancy. Trial moves that would displace disks across cell
boundaries are rejected. The order of the four checkerboard sub-sweeps
is also sampled as a random permutation. In this manner, an entire
sweep over $N$ particles satisfies detailed balance.  The reverse
sweep corresponds to an inverse shuffling and cell sequences with
opposite trial moves and occurs with equal probability.  To ensure
ergodicity, the cell system is randomly shifted after each sweep.  We
implement MPMC on a graphics processing unit (GPU) using CUDA. Details
are found in Ref.~\cite{Anderson_2012}.  The MPMC simulations execute
simultaneously on all 1536 cores of a NVIDIA GeForce GTX 680.

In event-driven molecular dynamics (EDMD), individual simulation
events correspond to collisions between pairs of disks
\cite{Alder_1957}.  The simulation is advanced sequentially from one
collision event to the next. Between collisions, disks move at
constant velocity (see \cite{Krauth_2006} for a basic discussion).
The computation of future collisions and the update of the event
schedule are performed efficiently using a binary tree and relating
searching schemes \cite{Rapaport_1980,Lubachevsky_1991,
  Marin_1993,Marin_1995}.  As a result, one collision event in EDMD
costs only about ten to twenty times more CPU time than a LMC trial
move, even for large system sizes. EDMD drives the system quite
efficiently through configuration space and clearly outperforms LMC.
The simulations with this algorithm use an Intel Xeon E5-1660 CPU with
a clock speed of 3.30GHz.

Event-chain Monte Carlo (ECMC)~\cite{Bernard_2009} replaces individual
trial moves by a chain of collective moves that all translate
particles in the same direction. At the beginning of each Monte Carlo
move, a random starting disk and a move direction are selected.  The
starting disk is displaced in the chosen direction until it collides
with another disk. This new disk is then displaced in the same
direction until another collision occurs or until the lengths of all
displacements add up to a total distance, an internal parameter of the
algorithm which is typically chosen such that the chain consists of $
\sim \sqrt{N}$ disks.  With periodic boundary conditions, ECMC is free
of rejections.  Global balance and ergodicity are preserved by moving
in two directions only, for example to the right and up. ECMC is
faster than LMC and EDMD~\cite{Bernard_2009}.  The simulations with
this algorithm were performed on an Intel Xeon E5620 CPU with a clock
speed of 2.40GHz.

For the large systems considered in this study, the ratio between the
large absolute particle coordinates and the potentially small
inter-particle distances becomes comparable to the accuracy of single
floating point precision, so that cancellation errors become critical.
Different strategies allow us to cope with this problem.  MPMC
performs all computations in single-precision because today's GPUs run
significantly slower in double-precision. We mitigate floating-point
cancellation errors by placing each particle in a coordinate system
local to its cell.  In this way, differences of relative positions are
less affected by floating point precision than absolute
positions. This strategy was also applied in~\cite{Bernard_2011}. EDMD
calculations are performed in double-precision to fully resolve
multiple coincident collisions and to span the entire time domain from
individual collision times to total simulation time.  The ECMC
algorithm is implemented for this work in double-precision. We use
this implementation to derive high-precision numbers at the density
$\eta=0.698$. Data points for ECMC at other densities are taken from
Ref.~\cite{Bernard_2011}, which employs single-precision. The
comparison of single-precision and double-precision calculations at
$\eta=0.698$ indicates that the two versions of ECMC yield the same
result for the pressure.

\subsection{Pressure computation} 

The complete statistical behavior of hard disks is contained in the
equation of state (pressure vs.\ volume or density), which requires
the precise evaluation of the internal pressure of the system.  The
equation of state allows computing the interfacial free energy and
tracking the changes in the geometry of coexisting phase regions in a
finite system (see \cite{Mayer_1965,Furukawa_1982,
  Schrader_2009,Bernard_2011}).

In the $NVT$ ensemble, the pressure is a dependent observable. In
Monte Carlo, it has to be computed from static configurations while,
in EDMD, it may in addition be derived from the collision rate. The
disparity of our approaches to calculate pressure constitutes one more
check for the implementations of our algorithms.

\subsubsection{Pressure from static configurations} 

In systems of isotropic particles with pairwise interactions, the
pressure can be computed from static configurations through the
pair-correlation function $g(r)$ \cite{Metropolis_1953}. The function
$g(r)$ is defined as the distribution of particle pairs at distance $r
= |\vec{x}_i - \vec{x}_j|$, normalized such that $g(r \to \infty)=1$.
In practice, particle distances are binned into a histogram with bin
size $\delta r$.  If $n$ out of $p$ pair distances are found to lie in
the interval $[r-\delta r/2, r+\delta r/2]$, then we have
\begin{equation}
  g(r) = \frac{n/p}{2 \pi r \delta r/V}.
\end{equation}

The pressure $P = -(\partial F / \partial V)_{T,N}$ is calculated from
the free energy $F = -\beta^{-1} \ln Z $ and the partition function
\begin{align}
  Z & = \frac{1}{N!} \int_0^L \dots \int_0^L dx_1 ....dx_{2N} \theta(x_1 \dots x_{2N})\nonumber \\
    & = \frac{V^N} {N!} \int_0^1 \dots \int_0 ^1 d\alpha_1 \dots d\alpha_{2N} 
        \theta(\alpha_1 \dots \alpha_{2N}),
\end{align}
where $\alpha_j = x_j/L$ are the particle coordinates relative to the
simulation box.  The Boltzmann weight $\theta$ is the characteristic
function for overlap, \textit{i.e.}\ zero if the configuration
contains overlaps and one otherwise.  A change of volume leaves the
$\alpha$ unchanged, but rescales the positions and the pair
distances. $\theta(\alpha_1 \dots \alpha_{2N})$ is only affected if
one of the pair distances is at contact, hence
\begin{align}
  \beta P &= \frac{N}{V} + \frac{\sigma}{V} \left.
             \mean{ \frac{\partial \theta}{\partial r}}
             \right\vert_{r = 2\sigma^{+}} \nonumber\\
          &= \frac{N}{V}(1 + 2 \eta\, g(2 \sigma^{+})).
\end{align}

\begin{figure}
  \includegraphics{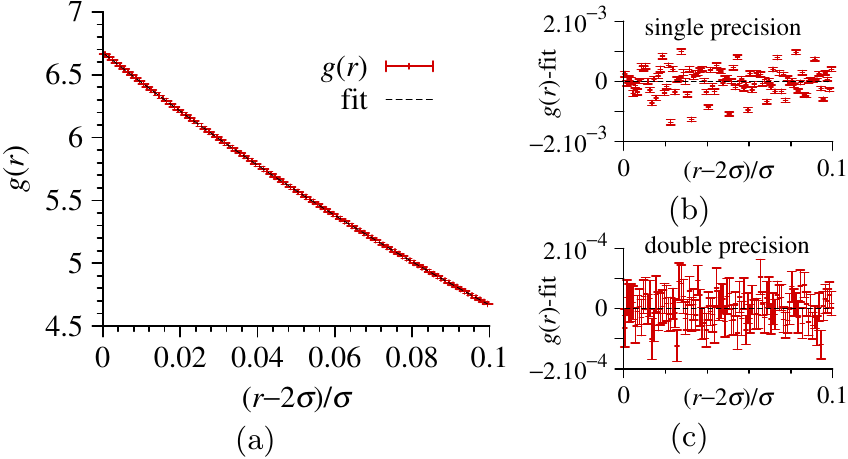}
  \caption
    {(Color online) (a)~Pair-correlation function $g(r)$ close to contact for
      $N=512^2$ at density $\eta = 0.698$ using LMC. Error bars are
      computed through $64$ independent simulations. $g(r)$ is fitted
      with a fourth-order polynomial.  Difference between $g(r)$ and
      the polynomial fit with the single-precision data~(b) and
      double-precision data~(c) for the histogram.}  \flabel{g_r}
\end{figure}

To access the contact value of $g(r)$, we fit the histogram of pair
distances obtained from LMC by a polynomial as shown in \fig{g_r} and
then extrapolate the fit to $r = 2 \sigma$ from the right. We choose
the bin size as $\delta r = 10^{-3} \sigma$. The histogram is limited
to $r \in [2\sigma, 2.1\sigma]$, and the fit is performed with a
fourth-order polynomial. These parameters are sufficient to obtain a
relative systematic error of less than $10^{-5}$. In the
single-precision version of the algorithm, $g(r)$ shows correlated
fluctuations which lead to systematic errors in the value of $g(r)$
for any single bin (see \fig{g_r}(b,c)). These errors are only due to
floating point round-off of the pair-correlation function (the sampled
configurations are essentially the same). However, they are periodic
with zero mean and do not affect the fit significantly. We verified
that single-precision rounding induces a relative systematic error
smaller than $10^{-5}$ on the estimated value of $g(2\sigma^{+})$.

\subsubsection{Dynamic pressure computation}

In molecular dynamics, static configurations can be analyzed as
before, but the pressure is computed directly and more efficiently
from the collision rate via the virial
theorem~\cite{Erpenbeck_1977,Alder_1960,Hoover_1967}. This avoids
binning and extrapolations.  The non-dimensional virial pressure in
two dimensions is given by
\begin{equation}
  \beta P = \frac{N}{V} \left[ 1-\frac{\beta m}{2 t_\text{tot}}\frac{1}{N}
            \sum_{\text{collisions} } b_{ij} \right]
\end{equation} 
where $t_\text{tot}$ is the total simulation time. The collision force
$b_{ij} = \vec{r}_{ij}\cdot \vec{v}_{ij}$ is defined between the
relative positions and relative velocities of the collision partners.
In equilibrium, the average virial for hard disks
equals~\cite{Ladd_1987}
\begin{equation}
  \mean{b_{ij}} = -2 \sigma \sqrt{\frac{\pi}{\beta m}}.
\end{equation}
Therefore, the pressure is simply given by the collision rate
$\Lambda=1/t_0$, the reciprocal of the mean free time $t_0$, as
\begin{equation}
  \beta P =  \frac{N}{V} \left[ 1 + \frac{\sigma \sqrt{\pi \beta m}}{2} \Lambda \right].
\end{equation}

To test the pressure computations, we compute the pressure with all
algorithms at one representative state point. With each algorithm, we
perform between 8 and 100 independent runs to compute the statistical
standard error. Results obtained with the four algorithms agree within
numerical accuracy to $\le 10^{-4}$, which is sufficient for our
purposes (see \tab{special_point}).

\begin{table}
  \tabcolsep1.5mm \centering
  \begin{tabular}{l c c c c}
    \hline\hline
    Algorithm & runs & disp./run & $\beta P(2\sigma)^2$ & std.\ error \\
    \hline
    LMC       & $64$   & $6 \times 10^{11}$ & $9.17046$ & $1.5 \times 10^{-4}$ \\
    EDMD      & $100$  & $10^{10}$          & $9.17076$ & $1.8 \times 10^{-4}$ \\
    ECMC      & $32$   & $5 \times 10^{11}$ & $9.17062$ & $8.7 \times 10^{-5}$ \\
    MPMC      & $8$    & $6 \times 10^{13}$ & $9.17078$ & $4.5 \times 10^{-5}$ \\
    \hline\hline
  \end{tabular}
  \caption{Test of the pressure computations for $N=256^2$ at $\eta = 0. 698$.
    The table lists for each algorithm the number of runs, number of displacements (disp.) per run, pressure, and standard error.
    Results of all four algorithms agree within their numerical accuracies.}
  \tlabel{special_point}
\end{table}

\section{Performance comparison}

One of the slowest processes during the time evolution of the
hard-disk system at high density is the fluctuation of the global
orientation order parameter
\begin{equation*}
  \Psi_6 = \frac{1}{N} \sum_j \psi_j,
\end{equation*}
which is the spatial average of the local orientational order parameter
\begin{equation}
  \psi_j = \frac{1}{6} \sum_{k= 1}^{6}  \exp(i\,6 \phi_{j,k}).
\end{equation}
The sum is over the six closest neighbors $k$ of disk $j$, and
$\phi_{j,k}$ is the angle between the shortest periodic vector
equivalent to $\vec{x}_k - \vec{x}_j$ and a chosen fixed reference
vector. This approach is simpler than using the Voronoi
construction~\cite{Bernard_2011}, without affecting the
autocorrelation functions. To determine the efficiency of our
algorithms, we track the autocorrelation function of
$\Psi_6$~\cite{Bernard_2009},
\begin{equation}
  C(\Delta t) = \frac{\mean{\Psi_6(t) \Psi_6^*(t+
                \Delta t)}_t}{\mean{|\Psi_6|^2}}.
\end{equation}
The $\Psi_6 \to \Psi_6+\pi$ symmetry in the square box imposes that
$C(\Delta t)$ decays to zero for infinite times. In the asymptotic
limit, the decay is exponential, $C(\Delta t)\propto\exp(-\Delta
t/\tau)$, and we obtain the correlation time $\tau$ from a fit of the
pure exponential part (see \fig{ACF}).

\begin{figure}
  \includegraphics{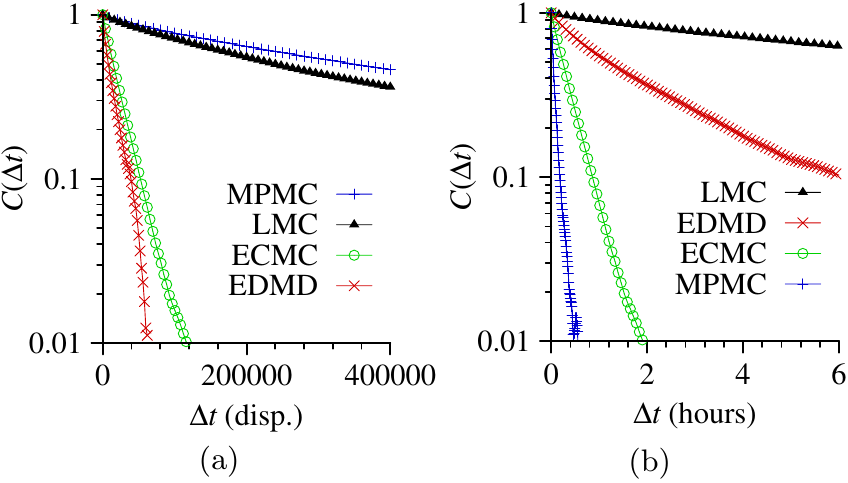}
  \caption
     {(Color online) Autocorrelation function of the global orientation order
       parameter $\Psi_6(t)$ for $N=512^2$, $\eta=0.698$ obtained with
       LMC, EDMD, ECMC, and MPMC. (a)~Time is measured in number of
       attempted displacements (or collisions) per disk. (b)~Time is
       measured in CPU or GPU hours.}  \flabel{ACF}
\end{figure}

We compare speeds for $N=512^2$ at $\eta= 0.698$, that is, in the
dense liquid close to the liquid--hexatic coexistence. Although still
a liquid, the correlation length of the local orientational order
parameter $\psi_j$ at this density is $\sim 50 \sigma$.  Such a large
correlation {\it length} induces a long correlation {\it time}, and
equilibration requires $>10^6$ trial moves per disk for LMC.  For the
test, each algorithm is set to its optimal internal parameters. Total
simulation run times are on the order of $10^3\tau$ to $10^4\tau$.

\fig{ACF} illustrates that the autocorrelation functions decay roughly
as pure exponentials. The time unit in \fig{ACF}(a) corresponds to the
number of attempted displacements per disk (number of collisions in
the case of EDMD and ECMC). As expected, MPMC decays slightly more
slowly than LMC because trial moves across cell boundaries are
rejected.  ECMC and EDMD are significantly faster than LMC, confirming
that these two methods sample configuration space more efficiently,
but slower than MPMC.  The time unit in \fig{ACF}(b) corresponds to
the real simulation time (CPU or GPU time).

Correlation times, number of attempted displacements per hour, and
accelerations with respect to LMC are summarized in \tab{speed}.  EDMD
and ECMC sample configuration space more efficiently by a factor of 39
times and 28 times, respectively, while MPMC samples configuration
space slightly less efficiently by a factor of 0.9. Evidently, the
efficiency of the simple LMC algorithm is improved significantly with
speed-ups of $10$, $70$, and $320$ for EDMD, ECMC, and MPMC,
respectively. Although speeds in \tab{speed} correspond to somewhat
different hardware, as indicated in the methods section, the numbers
give a clear idea of practical improvements that can be obtained with
respect to LMC.

\begin{table}
  \tabcolsep1.5mm
  \centering
  \begin{tabular}{l c c c c}
    \hline \hline
    Algorithm & $\tau$/disp. & disp./hour  & $\tau_{\text{LMC}}/\tau$
    & Speed-up \\
    \hline
    LMC       & $7\times 10^5$   & $6.5\times 10^9$    & $1$   & $1$   \\        
    EDMD      & $1.8\times 10^4$ & $1.7\times 10^9$    & $39$  & $10$  \\
    ECMC      & $2.5\times 10^4$ & $1.6\times 10^{10}$ & $28$  & $70$  \\
    MPMC      & $8\times 10^5$   & $2.3\times 10^{12}$ & $0.9$ & $320$ \\
    \hline \hline
  \end{tabular}
  \caption{Speed comparison of the four hard-disk algorithms for $N =
    512^2$, $\eta = 0.698$. The correlation time $\tau$ is measured in
    number of displacements (or collisions) per disk. disp./hour represents
    the number of displacements per hour achieved in
    our implementations. The two rightmost columns show the speed-up
    of the algorithms in number of displaced disks, and in terms of
    CPU or GPU time in comparison to LMC.}  \tlabel{speed}
\end{table}

\section{Results and discussion}

\subsection{Equation of state at high density}

In their seminal work, Alder and Wainwright observed a loop in the
equation of state of hard disks~\cite{Alder_1962}. As explained by
Mayer and Wood~\cite{Mayer_1965} (see also \cite{Furukawa_1982,
  Schrader_2009}), this loop is a result of finite simulation sizes
and therefore differs conceptually from a classic van der Waals loop,
which is derived in the thermodynamic limit.  The branches of the
Mayer-Wood loop are thermodynamically stable, but vanish in the limit
of infinite size.

It is known that the presence of a Mayer-Wood loop in the equation of
state is observed in systems showing a first-order transition as well
as systems showing a continuous
transition~\cite{Alonso_1999}. However, the behavior of these loops
with increasing system size is different. For a first-order
transition, the loop is present in the coexistence region and is
caused by the interface free energy $\Delta F$.  At a given density,
the interface free energy per disk, $\Delta f=\Delta F / N$, can be
computed by integrating the equation of state~\cite{Bernard_2011}. In
two dimensions, it scales as $\Delta f \propto N^{-1/2}$. In contrast,
for a continuous transition, $\Delta f$ decays faster, normally such
that $\Delta F$ is constant, that is $\Delta f \propto N^{-1}$, and
the equation of state becomes monotonic for large enough systems. The
scaling of $\Delta f$ with system size, together with a fixed finite
separation of the peaks for large system sizes, is a reliable
indicator of the first-order character of a phase
transition~\cite{Lee_1991}.

\begin{figure}
  \includegraphics{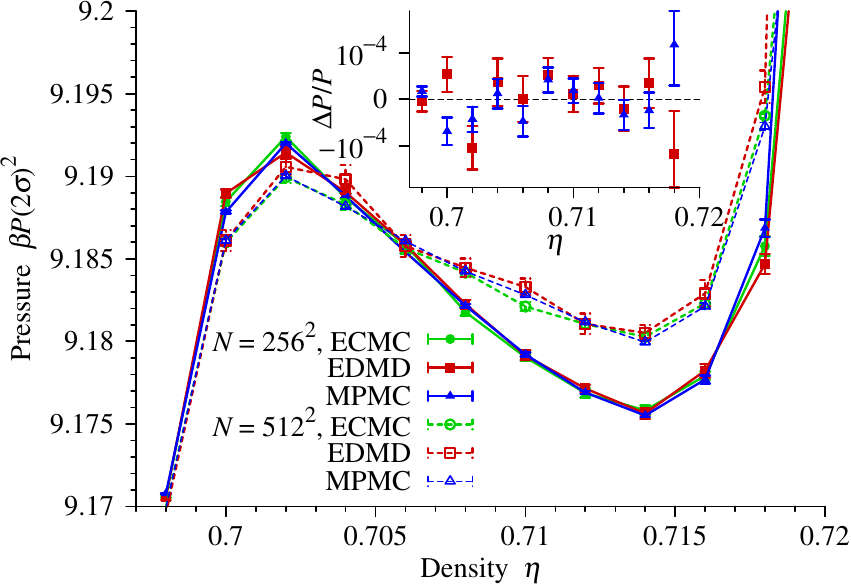}
  \caption
    {(Color online) Equation of state from ECMC, EDMD, and MPMC for $N=256^2$ and
      $N=512^2$. Error bars are mostly smaller than the
      symbols. Results agree within one standard deviation. The inset
      shows the relative pressure difference $\Delta P/P$ of EDMD and
      MPMC with respect to ECMC for $N=256^2$.}  \flabel{eqofstate}
\end{figure}

\begin{figure}
  \includegraphics{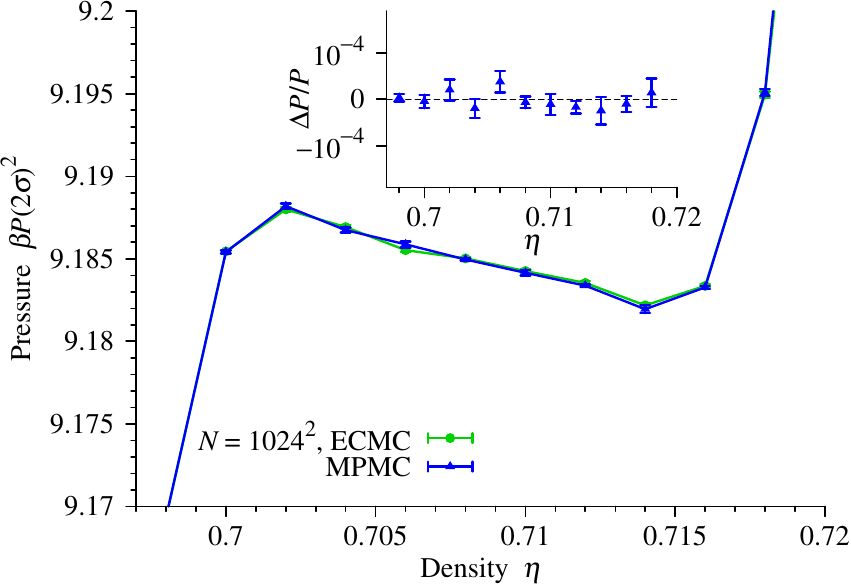}
  \caption
    {(Color online) Equation of state from ECMC and MPMC for $N=1024^2$.  Error bars
      are mostly smaller than the symbols.  Results agree within one
      standard deviation. The inset shows the relative pressure
      difference $\Delta P/P$ of MPMC with respect to ECMC.}
    \flabel{eqofstate_1024}
\end{figure}

\fig{eqofstate} shows the equation of state for $N=256^2$ and
$N=512^2$, obtained with ECMC, EDMD, and MDMC.  Error bars (standard
errors) are computed through independent simulations. The inset shows
the relative pressure difference $\Delta P/P$ of EDMD and MDMC with
ECMC. Error bars correspond again to the standard error on $\Delta
P/P$. We observe that the three independent simulations agree.  In a
similar way, \fig{eqofstate_1024} compares results from ECMC and MPMC
for $N=1024^2$. For this system size and the currently available
computer hardware, equilibration with EDMD takes too long to be
practical. Again, all results agree within standard deviations. Note
that while the Mayer-Wood loop shrinks with system size, the position
of the local extrema stay fixed close to $\eta=0.702$ and
$\eta=0.714$.

\subsection{Orientational order parameter field}

The degree and distribution of local order in a system can be analyzed
with the help of order parameters. By averaging a given order
parameter attached to each particle over a small sampling area
surrounding the particle, we obtain a continuous function, the
corresponding order parameter field.  Typical sampling areas used in
this work contain between 5 and 20 disks.

To show the separation of the liquid and the hexatic phase, we
graphically represent the orientational order parameter field
$\psi(\vec{x})$ for configurations at densities where coexistence
occurs. As for any first-order phase transition in two dimensions, the
characteristic geometry of the region of the minority phase with
increasing density is expected to change from an approximately
circular bubble into a parallel stripe and again into a circular
bubble.  Ref.~\cite{Bernard_2011} used the projection of the local
orientational order $\psi_k$ on the global orientational order
$\Psi_6$ and a linear color code. This projection is not a unique
measure for the orientational order parameter field. Instead, in
\fig{LocOrderSnapshot}, we use a circular color code on data obtained
by long MPMC simulations with $N=1024^2$.

We observe that the system is uniformly liquid at $\eta=0.700$ and the
color fluctuates on the scale of the correlation length. At
$\eta=0.704$, the representations indicate the presence of a circular
bubble of hexatic phase, visible as a large region of constant purple
color, whereas at $\eta=0.708$ a stripe minimizes the interfacial free
energy. The hexatic phase is now visible in blue, while the liquid is
characterized by fluctuations of the direction of $\Psi_6$.  The
constant color in the region of the hexatic phase confirms the
presence of the same orientational order across the system.
Additional important evidence for the nature of the transition is
provided by the spatial correspondence between variations in local
density and orientational order, which are included as a movie in the
supplemental material of this work. The movie illustrates for the
density $\eta=0.71$ that the system is ergodic by seeing the patches
of the two phases appear and disappear at different locations but with
roughly fixed ratio of areas.

\begin{figure}
  \includegraphics{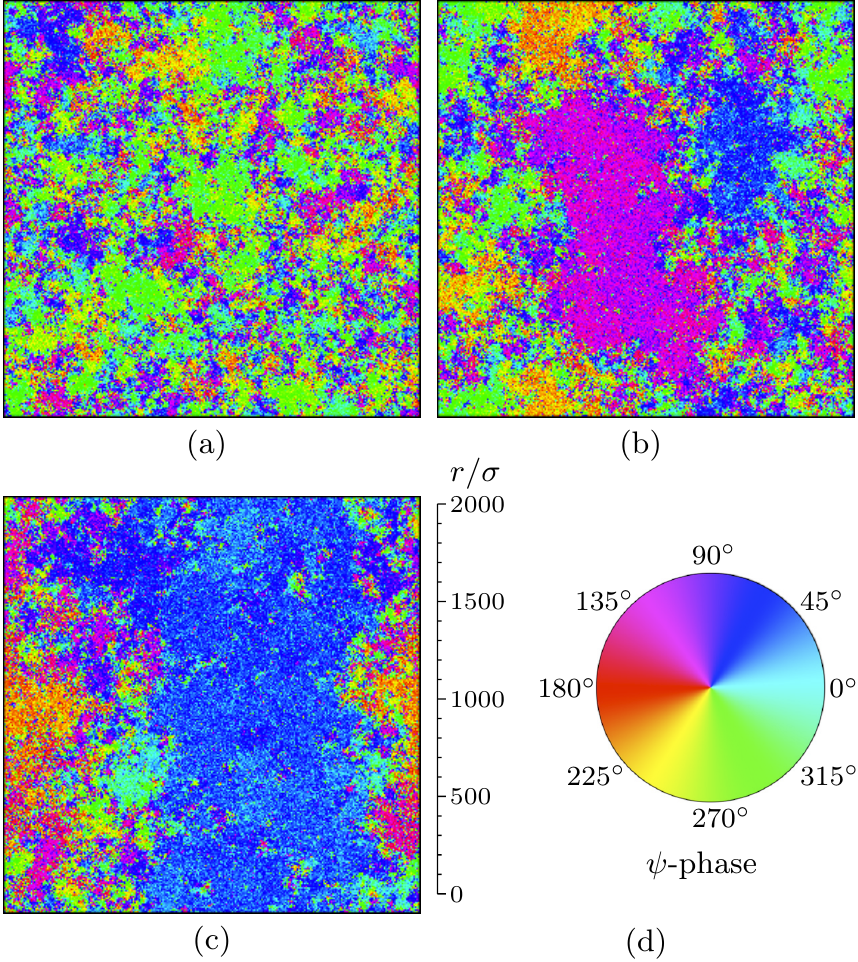}
  \caption
    {(Color online) Orientational order parameter field $\psi(\vec{x})$ of
      configurations obtained with the MPMC algorithm for system size
      $N = 1024^2$. With increasing density, (a)~pure liquid ($\eta =
      0.700$), (b)~a bubble of hexatic phase ($\eta = 0.704$), and (c)~a
      stripe regime of hexatic phase ($\eta = 0.708$) are visible. The
      interface between the liquid and the hexatic phase is extremely
      rough. (d)~A scale bar illustrates the size of the fluctuations. The phase of $\psi$
      is represented \emph{via} the color wheel.}
    \flabel{LocOrderSnapshot}
\end{figure}

\subsection{Positional order parameter field}

To identify the structure of the ordered phase in coexistence with the
liquid, we analyze the positional order in the system. The goal is to
distinguish the hexatic phase, which has short-range positional order
(characterized by exponential decay of the correlation function) and
quasi-long-range orientational order (algebraic decay), from the
two-dimensional solid, which has quasi-long range positional
order~\cite{Mermin_1968} and long-range orientational order (no
complete decay).

\begin{figure}
  \includegraphics{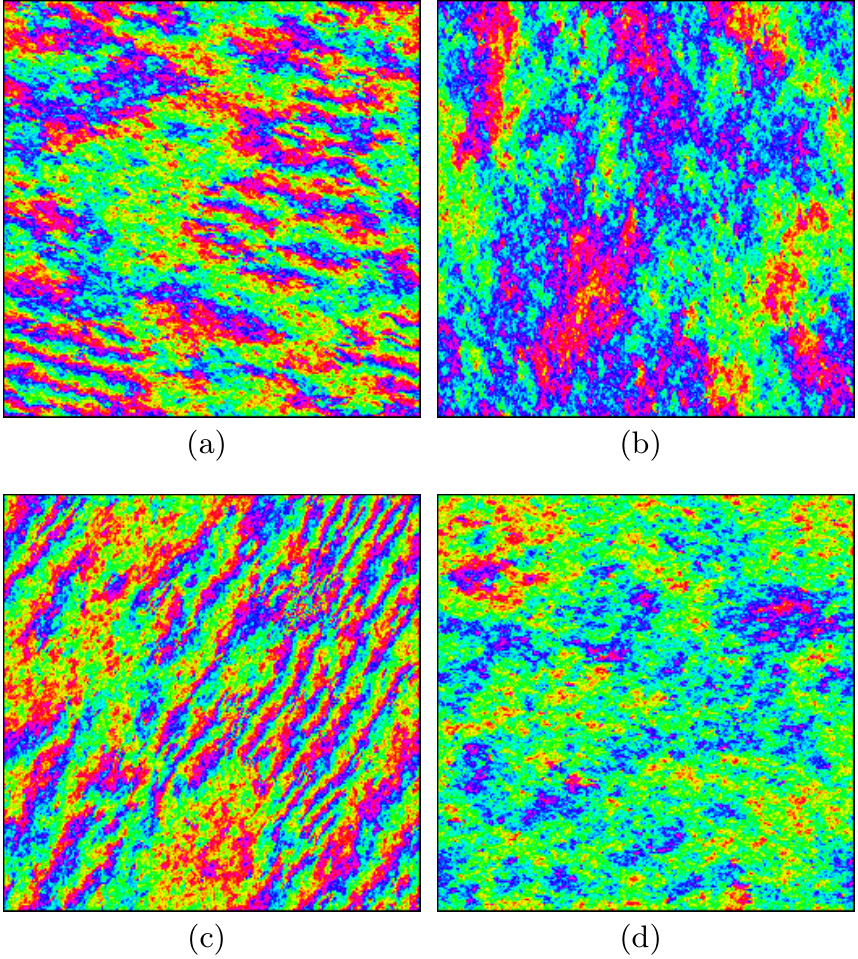}
  \caption
    {(Color online) Positional order parameter field $\chi(\vec{x})$ of
      configurations obtained with (a,b)~the MPMC and (c,d)~ECMC algorithms for
      system size $N = 1024^2$. (a,c)~In the hexatic phase ($\eta = 0.718$),
      positional order is short-range. (b,d)~Towards higher density
      ($\eta=0.720$), fluctuations are much weaker and bounded, as
      expected for a continuous transition to a solid phase. The scale
      bar and the color code for the $\chi$-phase are identical to the
      scale bar and the color code in \fig{LocOrderSnapshot}.}
    \flabel{PosOrderSnapshot}
\end{figure}

In~\cite{Bernard_2011}, the positional order was analyzed using the
two-dimensional pair-correlation function in direct space and the
decay of the positional correlation function at the wave vector
$\vec{q}_0$ corresponding to the maximum value of the first
diffraction peak of the structure factor
\begin{equation}
  S(\vec{q}) = \frac{1}{N} \sum_{n,m} \exp(i\vec{q} \cdot (\vec{x}_n -
\vec{x}_m)).
\end{equation}
With this classic method, the wave vector $\vec{q}_0$ must be chosen
carefully to correspond to a diffraction peak. In some previous
works~\cite{Mak_2006,Bagchi_1996}, it was assumed that $\vec{q}_0$
would correspond to the reciprocal vector of a perfect triangular
lattice of edge length $a_0$, namely $|\vec{q}_0|=2 \pi/(a_0
\sqrt{3}/2)$.  This assumption is not correct because the solid phase
has a finite density of vacancies and other defects in equilibrium,
which increases the effective lattice constant~\cite{Bernard_2011}.

We visualize the positional order parameter field $\chi(\vec{x})$
calculated from the positional order parameter
\begin{equation}
  \chi_j = \exp(i\vec{q}_0 \cdot \vec{x}_j).
\end{equation}
slightly above the upper critical density of coexistence with the
liquid~(\fig{PosOrderSnapshot}). At $\eta=0.718$, the system resembles
a patchwork of independent, solid-like regions of size in the order of
a few hundred $\sigma$.  Some regions show almost constant $\chi$,
while others are characterized by regular interference fringes
(oscillatory waves) with a fixed wave vector.  It can be shown that
each end of a fringe corresponds to one unpaired
dislocation~\footnote{Lattice fringe images have been frequently used
  in electron microscopy to analyze the positions and Burgers vectors
  of dislocations in diffraction contrast~\cite{Howie_1962}. Each
  discontinuity in the lattice fringe image, \emph{i.e.}\ an end of a
  fringe, corresponds to the presence of a dislocation at this
  position.}. We find that fringes can be made to disappear separately
in each region by small rotations of $\vec{q}_0$ around the
origin. This behavior is consistent with the existence of small-angle
grain boundaries separating neighboring solid-like regions, exactly as predicted
by the KTHNY scenario~\cite{Fisher_1979}.  Note that if fringes
disappear on one side of a boundary separating two regions, they
necessarily have to reappear on the other side with the sequence of
the colors reversed.

On physical grounds, the hexatic phase is not expected to be stable up
to close packing. Indeed, already at $\eta=0.720$, the positional
order field shown in \fig{PosOrderSnapshot} fluctuates much more
slowly and is highly correlated throughout the system as expected for
a solid phase.

\subsection{Positional correlation function}

The decay of positional order can be analyzed using the positional
correlation function in reciprocal space,
\begin{equation}
  C_{\vec{q}_0}(r) = \mean{\exp(i\vec{q}_0 \cdot (\vec{x}_n - \vec{x}_m))}.
\end{equation}
The averaging for $C_{\vec{q}_0}(r)$ is done on two levels. First, we
average over neighboring pairs that satisfy $|\vec{x}_n - \vec{x}_m|
\in [r-\sigma,r+\sigma]$. In addition, we conduct an average over
independent configurations, which can be a time average or an ensemble
average. For details on the configuration averaging see
Appendix~\ref{configurational_averaging}. As the system can perform global rotations during the
simulations, $\vec{q}_0$ rotates from one configuration to the
other. Each configuration is individually rotated so that $\Psi_6$ is
aligned in the same direction for all of them.  We verified that
alignment errors are sufficiently small and can be neglected.  As a
result of the configuration average, finite-size effects present at large
distances $r$ in the form of interferences are suppressed.

\fig{positional_order} shows $C_{\vec{q}_0}(r)$ at $\eta=0.718$ and
$\eta=0.720$. The results of ECMC and MPMC are again in good
agreement. We observe that $C_{\vec{q}_0}(r)$ decays exponentially at
$\eta=0.718$. Therefore, $\eta=0.718$ cannot be in the solid phase.
The length scale of the exponential decay is in the order of
$100\sigma$, which corresponds approximately to the width of the
interference fringes in \fig{PosOrderSnapshot}.  Since the coexistence
phase ends at $\eta \simeq 0.716$, the region $\eta \gtrsim 0.716$ is
thus hexatic. We have shown once more that the first-order transition
observed in \fig{eqofstate} connects a liquid and a hexatic phase.

\begin{figure}
  \includegraphics{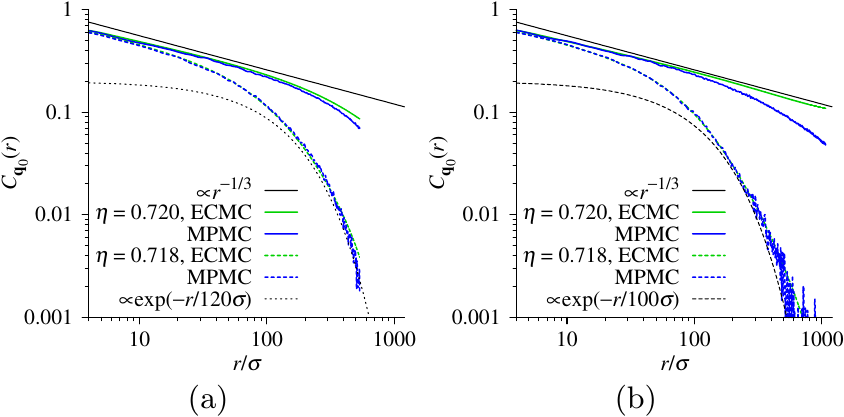}
  \caption
     {(Color online) The positional correlation function $C_{\vec{q}_0}(r)$ shows
       exponential decay at density $\eta=0.718$ and approaches a
       power law $\propto r^{-1/3}$ at $\eta=0.720$. (a)~System size $N=512^2$: Excellent
       agreement between ECMC and MPMC. (b)~System size $N=1024^2$: Excellent agreement in the
       hexatic phase ($\eta = 0.718$) and fair agreement at the
       approach of the solid phase ($\eta = 0.720$). Our algorithms
       fall out of strict equilibrium in the solid and long-scale
       correlations become sensitive to the boundary conditions.}
     \flabel{positional_order}
\end{figure}

The positional order increases drastically at $\eta=0.720$.
$C_{\vec{q}_0}(r)$ decays almost as a power law, $r^{-1/3}$, which is
the stability limit for the solid phase in the KTHNY theory. Thus, the
stability regime of the hexatic phase comprises a narrow range of
density. An additional characterization of the hexatic phase with ECMC
including a study of the diffraction peak shape can be found
in~\cite{Bernard_2011b} and in the supplemental material
of~\cite{Bernard_2011}. The continuous transformation to the solid
phase and the nature of the hexatic phase agree with the KTHNY
scenario.

Slight variations in the positional correlations can be observed in
\fig{positional_order} at density $\eta=0.720$ for distances
comparable to the system size.  Two factors play a role. First, the
relaxation becomes very slow at the onset of the solid phase. Our
largest system no longer achieves full global rotations with respect
to $\Psi_6$. Second, the positional correlations span the whole
simulation box and therefore depend slightly on the orientation of the
crystal. Longer and larger simulations are necessary to determine the
location of the hexatic--solid transition with good
precision. However, the general absence of a loop in pressure is
sufficient to rule out a first-order transition. We note that while
our simulations fall out of strict equilibrium with respect to global
rotations in the solid phase, they remain fully ergodic within our
simulation times on both sides of the liquid--hexatic transition.

\section{Conclusion}

We analyzed the thermodynamic behavior of the hard-disk system close
to the melting transition using independent implementations of three
different simulation algorithms to sample configuration space and two
distinct approaches for the pressure computation.  The equation of
state data of Ref.~\cite{Bernard_2011} are confirmed within numerical
accuracy both qualitatively and quantitatively. Typical relative
errors are $\lesssim 10^{-4}$, more than one order of magnitude
smaller than finite-size effects for systems with up to $N=1024^2$
particles. Such finite-size effects are manifested in the form of a
Mayer-Wood loop in the equation of state.  Our analysis of
orientational and positional order parameters confirms the presence of
a first-order phase transition from liquid order to hexatic order and
a continuous phase transition from hexatic order to a solid phase.

\appendix
\section{Role of configuration averaging}
\label{configurational_averaging}

\begin{figure}
  \includegraphics{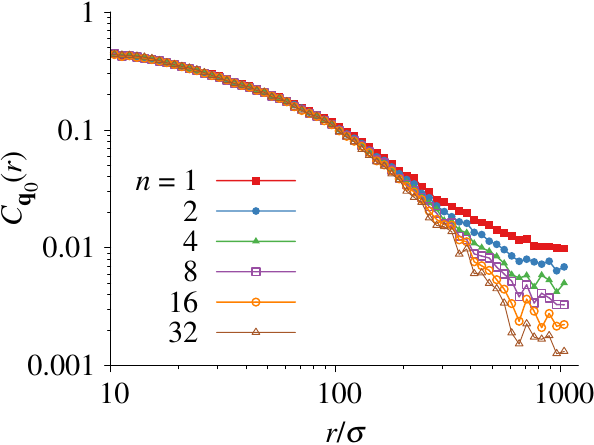}
  \caption
     {(Color online) Influence of the configuration averaging on the positional
       configuration functions in the hexatic phase. For single
       configurations ($n=1$, data incoherently averaged), positional
       correlations cannot decay below a level given by the square
       root of the ratio of the correlated domain size to the system
       size. Coherent averaging over $n=2$, $4$, $8$, $16$, and
       $32$ configurations reduces the noise level and the
       correlations. The data corresponds to a single long MPMC run
       with $N=1024^2$ particles at density $\eta=0.718$.}
       \flabel{positional_order_averaged}
\end{figure}
It is instructive to analyze the influence of configuration averaging
that we employ to calculate the correlation function
$C_{\vec{q}_0}(r)$.  We define a configuration average as the time
average or the ensemble average of $C_{\vec{q}_0}(r)$ over individual configurations.  This
coherent averaging procedure stands in contrast to the incoherent
averaging procedure, where we bin the function $C_{\vec{q}_0}(r)$,
determine the maximum within each of the bins, and average the maxima. 
As shown in \fig{positional_order_averaged},
for incoherently averaged single configurations the
long-distance correlations do not decay below a sampling threshold
that is set by the inverse square root of the number of independent
domains in the sample. For our large system of $N=1024^2$ particles,
given that the sample size is $L/2 \sim 1000\sigma$ and the positional
correlation length is in the order of $100 \sigma$, there are about
100 independent domains.  Residual correlations visible in the figure
as a plateau at large values of $r/\sigma$ correspond to what would be
expected from about $100$ independent randomly positioned (yet equally
oriented) lattices. As illustrated in the figure, coherent averaging
over oriented configurations increases the effective number
of independent domains, thus to reduce the noise level.  The same
effect would be obtained if we could equilibrate yet larger systems.

\begin{acknowledgments}
M.E., J.A.A., and S.C.G.\ acknowledge support by the Assistant
Secretary of Defense for Research and Engineering, U.S.\ Department of
Defense No.\ N00244-09-1-0062.  M.I.\ is grateful for financial
support from the CNRS-JSPS Researcher Exchange Program for staying at
ENS-Paris and Grant-in-Aid for Scientific Research from the Ministry
of Education, Culture, Sports, Science and Technology No.\ 23740293.
W.K.\ acknowledges the hospitality of the Aspen Center for Physics,
which is supported by the National Science Foundation Grant
No.\ PHY-1066293.  MPMC simulations were performed on a GPU cluster
hosted by the University of Michigan's Center for Advanced Computing.
EDMD simulations were partially performed using the facilities of the
Supercomputer Center, ISSP, University of Tokyo, and RCCS, Okazaki,
Japan.  We acknowledges helpful correspondence and discussions with
B.J.\ Alder, D.\ Fiocco, S.C.\ Kapfer, and S.\ Rice.
\end{acknowledgments}

\end{document}